\documentclass[12pt, layout=twocolumn]{iopart}
\usepackage{hyperref}
\usepackage{tikz,amssymb}
\expandafter\let\csname equation*\endcsname\relax
\expandafter\let\csname endequation*\endcsname\relax
\usepackage{amsmath}
\usepackage{algorithm,algpseudocode}
\usetikzlibrary{decorations.markings,angles,quotes}

\newcommand{\norm}[1]{\parallel #1 \parallel}
\def\v#1{\textbf{#1}} 

\begin{document}
    
    \title[]{A Methodology to Generate Crystal-based Molecular Structures for Atomistic Simulations}
    
    \author{Christian F. A. Negre$^1$, Andrew Alvarado$^{2,3}$, Himanshu Singh$^1$, Joshua Finkelstein$^1$, Enrique Martinez$^{3,4}$, and Romain Perriot$^1$}
    
    \address{1\ Theoretical Division, Los Alamos National Laboratory, Los Alamos, NM 87545, USA}
    \address{2\ Advances System Development, Los Alamos National Laboratory, Los Alamos, NM 87545, USA}
    \address{3\ Department of Mechanical Engineering, Clemson University, Clemson, SC 29623, USA }
    \address{4\ Department of Materials Science and Engineering, Clemson University, Clemson, SC 29623, USA }
    
    \ead{cnegre@lanl.gov}
    
    \vspace{10pt}
    \begin{indented}
        \item[] \today
    \end{indented}
    
    \begin{abstract}
        We propose a systematic method to construct crystal-based molecular structures often needed as input for computational chemistry studies.
        These structures include crystal ``slabs" with periodic boundary conditions (PBCs) and non-periodic solids such as Wulff structures. 
        We also introduce a method to build crystal slabs with orthogonal PBC vectors. 
        These methods are integrated into our code, \texttt{Los Alamos Crystal Cut} (LCC), which is open source and thus fully available to the community. 
        Examples showing the use of these methods are given throughout the manuscript.
    \end{abstract}
    
    %
    \vspace{2pc}
    \ioptwocol

    \section{Introduction}
    Surface science is essential to understand and predict many physical phenomena including heterogeneous catalysis \cite{Schmickler2010-al,Shetty2020-vg}, photo-catalysis \cite{BASAVARAJAPPA20207764,LIANG202089}, material interfaces \cite{Hakkinen2012,Batzill2012}, and optical properties \cite{Anfuso2012-bs,Wang2012}. Surface science is also crucial to study the shape and properties of nanocrystals~\cite{Boles2016}, which are essential to quantum dot applications~\cite{Robin2016, Martinez2018}, and 2D materials, exhibiting exceptional electrical, optical and mechanical properties~\cite{Novoselov2016,VahidMohammadi2021}. Beyond physical chemistry, surface science also plays a crucial role in biomedical~\cite{Castner2002} and bioengineering~\cite{Tirrell2002} applications, and dictates crystal growth~\cite{noauthor_1990-qo}, which is known to affect, for instance, the performance of high explosives~\cite{Setchell1984,Price1988}. Moreover, in all applications where the material exhibits a high surface to volume ratio, the properties of the surface (exposed crystal faces) largely determine the properties of the material.
    
    Despite the progress of characterization techniques, simulations remain a fundamental part of surface science, either to complement~\cite{Dahal2014} or fully predict~\cite{Negre2008-rm, Fuertes2013-ky} the properties of surfaces, interfaces, and nanoparticles. Electronic transport, optical properties, and even surface reconstructions can be a significant challenge for empirical models and, in order to perform these simulations, \textit{ab initio} level of theory is often required due to the complexity of the phenomena involved~\cite{HAMMER200071}. In most cases, the first step involved in these calculations will involve building a model crystal slab, which should obey the following constraints: the system must give us access to the surfaces of interest to the particular problem, it must be periodic in all other directions, and, in order to improve computational efficiency, it must be as small as possible (electronic structure calculations are usually performed on hundreds to a few thousands atoms, at most). This results in a crystal-based parallelepiped with planes that are not necessarily orthogonal to each other since, in the general case, the unit cell is triclinic (i.e. the lattice vectors are non-orthogonal to each other with differing lengths and angles to one another).
    
\begin{figure}
        \begin{center}
            \begin{tikzpicture}
                
                \coordinate (O) at (0, 0);
                \node[inner sep=0pt] (benzene) at (O)
                {\includegraphics[width=.25\textwidth]{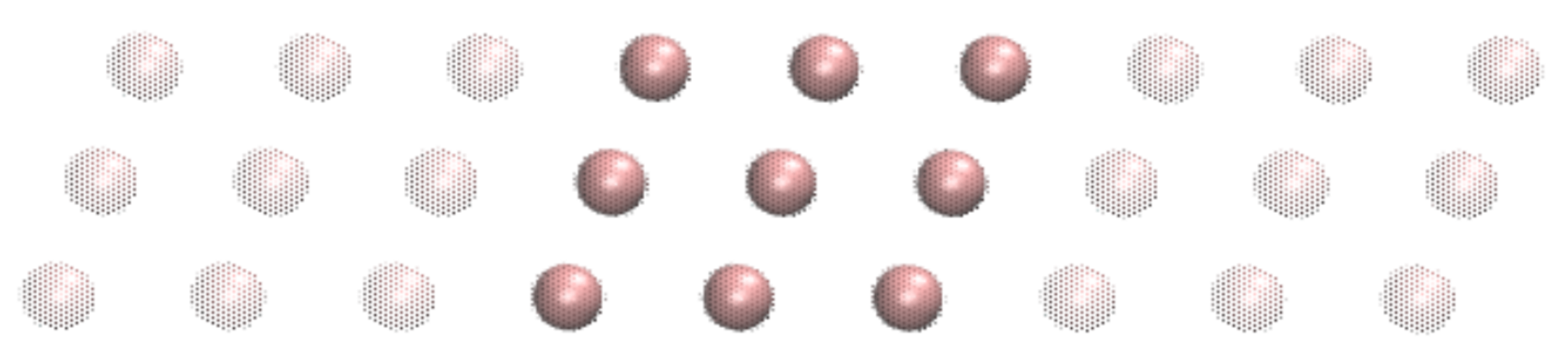}};
                \node[inner sep=0pt] (benzene) at (0.8,2)
                {\includegraphics[width=.25\textwidth]{idealSystem.pdf}};
                
                \draw[line width=2pt,red,-stealth](-0.6,-0.325)--(0.6,-0.325) node[anchor=north east]{$\v{p}_1$};
                \draw[line width=2pt,red,-stealth](-0.6,-0.3)--(0.175,1.6) node[anchor=north east]{$\v{p}_2$};
                \draw[line width=1pt,black,<-](0.0,0.5)--(0.5,1.0)--(2.0,1,0) node[anchor=west]{\scriptsize Surface of interest};
                
            \end{tikzpicture}
            \caption{Representation of a typical system slab needed to study a particular physical chemistry surface property. The system slab is composed of lattice points that are illustrated as bright magenta spheres together with the slab PBC vectors $\v{p}_1$ and $\v{p}_2$.}
            \label{idealSystem}
        \end{center}
    \end{figure}

    In order to study crystal surfaces with quantum chemistry methods, it is often necessary to have a crystal slab cut through planes that expose the face one wants to study and that also satisfies the periodic boundary conditions (PBCs) imposed by the crystal unit cell. A typical minimalistic system is depicted in Figure \ref{idealSystem}. %
    This is a $z-$axis view of a $3 \times 3 \times 3 = 27$ monoclinical unit cell system showing the slab PBC vectors. In this case the $\v{p}_2$ vector was enlarged in order to have some vacuum that could expose the surface of interest. 
    At first, this system may not seem complicated as it can be easily built with an \textit{ad-hoc} procedure using an off-the-shelf molecular visualization tool. There are however, many cases in which building such a system in this way could turn into a complicated and time consuming endeavor. Triclinic unit cells exposing some crystal face with large Miller indices fall within this category. Moreover, a lot of time and effort is consumed when errors in the simulations arise due to an ill-chosen system slab. How do we then proceed to construct any desired crystallographic system by just knowing the basic crystallographic data? In this article, we explain a method based on purely algebraic/geometrical transformations that leads to a sufficiently small crystal slab exposing the desired crystal faces. We would like to offer a detailed and simple step-by-step procedure that the reader could fully code up on their own. 
    Moreover, the method developed in this paper can also be used to construct Wulff type of structures provided that the exposed $(hkl)$ planes are known. 
    
    For many applications, it is preferable to build perfectly orthogonal faces to the exposed surface, i.e. orthorhombic systems. For instance, in shock simulations, a piston hits the back surface of the sample (or vice versa) and the shock propagates through the material oriented in a specific way~\cite{Perriot2012,Perriot2014b}; thermodynamic quantities are then estimated within slices of the material perpendicular to the shock direction, thus the use of orthogonal planes makes data processing a lot simpler. In addition, certain simulation codes explicitly require orthorhombic simulation boxes. 
    
    Previously, the authors of  Ref.~\cite{Kroonblawd2016-yb} proposed a method where the vectors defining the slab are tentatively constrained to satisfy both the orthogonality and periodicity conditions; however, in the general case, fulfilling these two conditions is not always possible. There is thus a balance between two effects: the larger the cell, the higher the probability of fulfilling PBCs, although resulting in an increase of the computational cost when the slab is used as an input for a quantum chemistry application code. On the other hand, a system that is not periodic will induce possibly large strain and stress, compromising some thermodynamic properties such as volume or pressure and yielding artificial responses. In Ref.~\cite{Kroonblawd2016-yb}, the algorithm usually produces a cell that is periodic but not exactly orthogonal, with small deviations in the lattice angles allowed to preserve this condition. This occasionally results in cells that do not have the exact requested orientation. In this paper we propose an efficient algorithm to build perfectly orthorhombic cells where the lattice periodicity mismatch is used to assess the validity of the slab.
    
    The following sections are organized as follows: We first introduce some basic crystallographic concepts in order to keep consistent notation throughout the manuscript. In Section \ref{extended} we introduce our method to cut a crystal lattice, and in Section \ref{extendedpbc} we develop the techniques to determine the PBC vectors. Section \ref{nonper} is dedicated to explaining how the method can be used to construct non-periodic solids using Wulff structures as an example. Finally, in Section \ref{ortho} we explain a method to construct crystal slabs with orthogonal PBC vectors. Sections in the Appendices are used for support and clarification throughout the text. Units of length and angles used in all the examples are in Angstroms (\AA) and degrees ($^{\circ}$) respectively. 
    
    \section{Background}
    \label{background}
    A crystal lattice is a set of points $\mathcal{L} \subseteq \mathbb{R}^3$ that is fully determined by the primitive unit cell described by the lattice vectors $\v{a}_1$, $\v{a}_2$, and $\v{a}_3$. For any point $\v{r}$ belonging to $\mathcal{L}$, there exist three integers, $n_1$, $n_2$, $n_3$ such that  
    \begin{equation}
        \v{r} = n_1 \v{a}_1 + n_2 \v{a}_2 + n_3 \v{a}_3 \;. 
        \label{translation}
    \end{equation}
    Formally, $\mathcal{L} = \{  \v{r}\in \mathbb{R}^3 \;|\; n_1, n_2, n_3 \in \mathbb{Z} \}$.
    A conventional unit cell (such as the cubical systems by Bravais), is just a more elaborate cell in which symmetry is increased. This increase of symmetry in some cases will, for instance, render lattice vectors that are orthogonal to each other; a highly desirable property for many applications. Regardless of which type of cell we have, the convention in crystallography is to report the so-called lattice 
    parameters $a$, $b$, $c$, $\alpha$, $\beta$, and $\gamma$; where $a$, $b$, and $c$ are the lengths of lattice vectors $\v{a}_1$, $\v{a}_2$, and $\v{a}_3$, respectively, and $\alpha$, $\beta$, $\gamma$ are, respectively, the angle between vectors  $\v{a}_2$ and $\v{a}_3$, $\v{a}_3$ and $\v{a}_1$, and $\v{a}_1$ and $\v{a}_2$ \cite{Ashcroft1976-uw}. This reduces the arbitrariness of having to choose a lattice orientation given by the lattice vectors. Note that if the lattice is rotated, our lattice vectors will need to be rotated as well, whereas the lattice parameters will stay the same.  
    Finally, a full representation of the system needs a ``basis,'' which is the minimal molecular fragment contained by each unit cell. It is common to express the coordinates of the basis in fractions of the lattice vectors. By choosing this coordinate system we make the orientation of the basis invariant to lattice rotations.
    
    Although working with lattice parameters has some advantages, it is convenient to compute the lattice vectors in order to do all the necessary transformations to build a PBC slab.
    In order to compute the lattice vectors from the lattice parameters one needs to apply the following transformations: 
    \begin{equation}
        \eqalign{
            \v{a}_{1x} &= a \\
            \v{a}_{1y} &= 0 \\
            \v{a}_{1z} &= 0       
        }\label{first_lattice_param}
    \end{equation}
    \begin{equation}
        \eqalign{
            \v{a}_{2x} &= b \cos(\gamma) \\
            \v{a}_{2y} &= b \sin(\gamma) \\
            \v{a}_{2z} &= 0  
        }
    \end{equation}
    \begin{equation}
        \eqalign{
            \v{a}_{3x} &=c \cos(\beta) \\
            \v{a}_{3y} &=c \frac{\left( \cos(\alpha)-\cos(\gamma) \cos(\beta) \right)} {\sin(\gamma)} \\
            \v{a}_{3z} &=\sqrt{\left(c^2 - \v{a}_{3x}^2 - \v{a}_{3y}^2\right)}
        }
    \end{equation}
    where we have arbitrarily set $\v{a}_1$ to be aligned with the $x$-axis, or in more formal terms, the first canonical vector $\v{e}_1 = (1,0,0)$ in the canonical basis for $\mathbb{R}^3$. An equivalent reverse transformation is used to compute the parameters given the lattice vectors:
    \begin{equation}
        \eqalign{
            a &= \sqrt{(\v{a}_{1x}^2+\v{a}_{1y}^2+\v{a}_{1z}^2)} \\
            b &= \sqrt{(\v{a}_{2x}^2+\v{a}_{2y}^2+\v{a}_{2z}^2)} \\
            c &= \sqrt{(\v{a}_{3x}^2+\v{a}_{3y}^2+\v{a}_{3z}^2)}
        }
    \end{equation}
    \begin{equation}
        \eqalign{
            \gamma &= \frac{360}{2 \pi} \arccos\left( (\v{a}_1 \cdot \v{a}_2)/(a b) \right) \\
            \beta &= \frac{360}{2 \pi} \arccos\left( (\v{a}_1 \cdot \v{a}_3)/(a c) \right) \\
            \alpha &= \frac{360}{2 \pi} \arccos\left( (\v{a}_2 \cdot \v{a}_3)/(b c) \right)
        } 
    \end{equation}
    
    \section{Building extended systems}
    \label{extended}
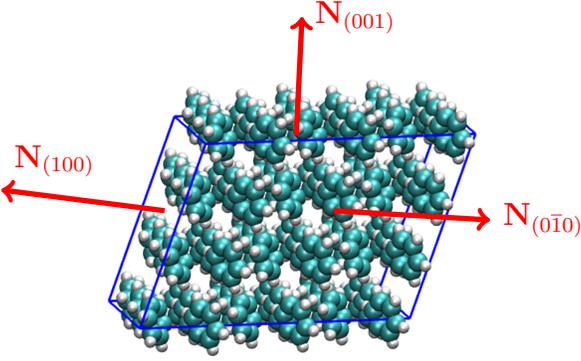
\begin{figure}
        \begin{tikzpicture}
            \coordinate (O) at (0, 0, 0);
            \node[inner sep=0pt] (benzene) at (O)
            {\includegraphics[width=.3\textwidth]{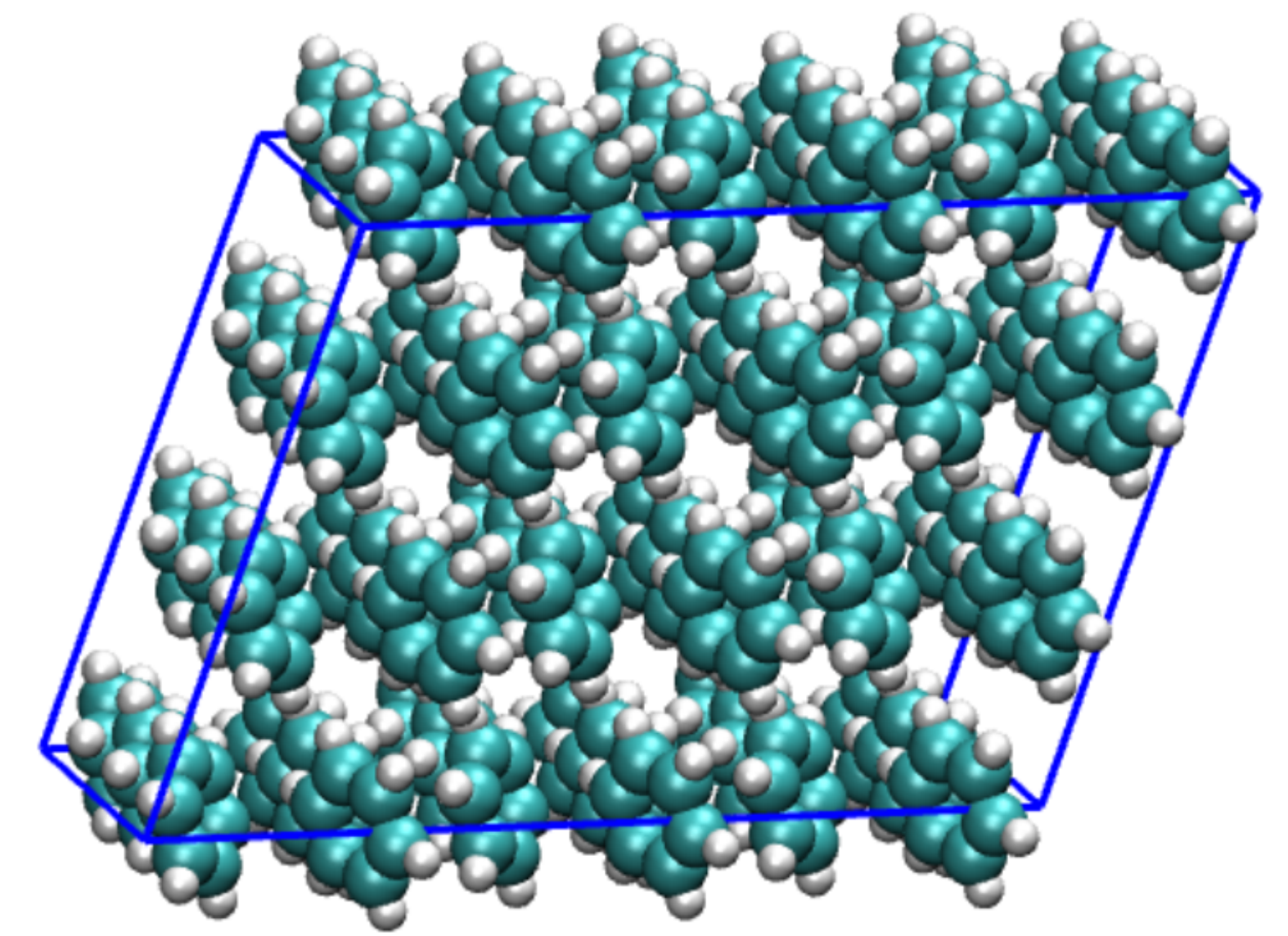}};
            \draw[line width=2pt,red,->] (0.0,1,-0.2) --++ (0.0,1.5,-0.2) node[anchor=west]{$\v{N}_{(0 0 1)}$};
            \draw[line width=2pt,red,->] (-1.75,0,-0.2) --++ (-2.25,0.2,-0.2) node[anchor=south west]{$\v{N}_{(1 0 0)}$}; 
            \draw[line width=2pt,red,->] (0.5,0,-0.2) --++ (2.0,-0.2,-0.2) node[anchor=west]{$\v{N}_{(0 \overline{1} 0)}$};     
        \end{tikzpicture}
        %
        
        \caption{Monoclinic structure of benzene. A $3 \times 3 \times 3$ slab showing exposed faces and plane normal vectors. Benzene molecules are represented with cyan and white spheres for every carbon and hydrogen atom, respectively. The normal vectors are orthogonal in the reciprocal basis but non-orthogonal in the canonical basis. Monoclinic benzene has $a$ = 5.5146, $b$ = 5.4951, 
            $c$ = 7.6536, $\alpha$ = $\gamma$ = 90.0, and $\beta$ = 110.6.}\label{benzeneSlab}
    \end{figure}

    Using the lattice vectors, a crystal slab can be built simply by adding lattice points according to Eq.~(\ref{translation}) for a finite number of $n_i$'s. 
    The resulting slab would expose the $(1 0 0)$, $ (0 1 0)$ and $(0 0 1)$ crystalline faces as well as the respective opposite faces given by $(\overline{1} 0 0)$, $(0 \overline{1} 0)$ and $(0 0 \overline{1})$. Note that this slab will form a parallelepiped whose edge directions are not necessarily orthogonal to one another in $\mathcal{E} \equiv \left\{\v{e}_1,\v{e}_2,\v{e}_3 \right\}$, the standard basis for $\mathbb{R}^3$. We shall call this slab the ``canonical slab.''
    An example canonical slab of the monoclinic phase of benzene is shown in Figure~\ref{benzeneSlab}. 
    At this point, a natural question emerges. What if now we need to expose other crystalline faces to perform specific computational physico-chemical studies? In this case, the periodicity of the slab will be key. The strategy we follow in this section is to cut out a crystal slab using Miller planes directly (planes given by a specific set of Miller indices) and then determine the PBC vectors for this slab: $\v{p}$, $\v{p}_1$, and $\v{p}_2$. The Miller planes are determined by the normal vectors perpendicular to the desired crystal faces. A Miller plane $(h k l)$  has normal vector $\v{N}_{(hkl)} = h \v{b}_1 + k \v{b}_2 + l \v{b}_3$, where the reciprocal lattice vectors $\v{b}_1$, $\v{b}_2$, $\v{b}_3$ are defined to be:
    \begin{equation}
        \eqalign{
            \v{b}_1 &\equiv \frac{2 \pi }{V} \v{a}_2 \times \v{a}_3 \\
            \v{b}_2 &\equiv \frac{2 \pi }{V} \v{a}_3 \times \v{a}_1 \\
            \v{b}_3 &\equiv \frac{2 \pi }{V} \v{a}_1 \times \v{a}_2 
        }
    \end{equation}
    where $V=|\v{a}_1 \cdot \v{a}_2 \times \v{a}_3|$ is the volume of the unit cell. 
    It is easy to see from the definition of the reciprocal vectors above that $\{\v{b}_1, \v{b}_2, \v{b}_3\}$ are biorthogonal to $\{\v{a}_1, \v{a}_2, \v{a}_3\}$, i.e.
    $\v{a}_i \cdot \v{b}_j = 0$ if $i \neq j$. 
    %
    
    \subsection{Cutting by planes}
    
    We now discuss a general algorithm to perform a cut by any plane in $\mathbb{R}^3$ and we then apply this to the particular case of Miller planes in order to build out the desired slab.
    
    As before, let $\mathcal{L}$ be the set of crystal lattice points and now let $\Pi = \{\v{r}\in \mathbb{R}^3 \; \vert \; (\v{r}-\v{Q})\cdot\v{N} = 0\}$ be the set of points $\v{r}$ defining a plane with normal vector $\v{N}$, passing through the point $\v{Q}$. Without loss of generality, and provided $\Pi$ does not intercept the origin, the center $\v{Q}$ can perfectly well be chosen to
    align with the normal vector $\v{N}$. In this sense,  $\Pi(c) =  \{\v{r}\in \mathbb{R}^3 \; \vert \;  (\v{r}-c\v{N})\cdot\v{N} = 0\}$ defines a set of parallel planes $\{\Pi(c) \}$ parameterized by $c$ all with normal vector $\v{N}$.
    We are now interested in selecting all the lattice points that are ``below'' the plane $\Pi(c)$. To do this we just need to evaluate the sign of the inner product between the vector $\v{r} - c\v{N}$ and the normal vector $\v{N}$. By setting $\v{n} = c\v{N}$, the condition for keeping a lattice point $\v{r}$ reads
    as $(\v{r}-\v{n})\cdot \v{N} < 0$, so that
    \begin{equation}
        \left\{\v{r} \in \mathcal{L} \; \vert \;  (\v{r}-\v{n})\cdot \v{N} < 0 \right\}
        \label{cut_with_N}
    \end{equation}
    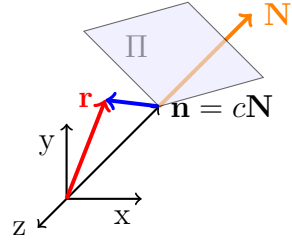
\begin{figure}
        \centering
        \begin{tikzpicture}
            \coordinate (O) at (0, 0, 0);
            \coordinate (A) at (2,2,2);
            \coordinate (B) at (4,4,4);
            \coordinate (posPi) at (1.7,2.8,2);
            \coordinate (C) at (0.9,1.7,1);    
            \draw[thick,->] (O) --++ (1,0,0) node[anchor=north east]{x};     
            \draw[thick,->] (O) --++ (0,1,0) node[anchor=north east]{y};     
            \draw[thick,->] (O) --++ (0,0,1) node[anchor=east]{z};
            \draw[thick,->] (O)--(A) node[anchor=west]{$\v{n} = c \v{N}$};
            \draw[thick,->,orange,,line width=1.5pt] (A)--(B) node[anchor=west]{$\v{N}$};
            \draw[->,red,line width=1.5pt] (O)--(C) node[anchor=east,red]{$\v{r}$};
            \draw[->,blue,line width=1.5pt] (A)--(C) node[anchor=west,red]{};     
            \node[inner sep=0pt] (pi) at (posPi){$\Pi$};
            \filldraw[fill=blue!10, opacity=0.6] (2,2,2) -- (3,2,1) --  (2,3,1) -- (1,3,2) -- (2,2,2);
        \end{tikzpicture}
        \caption{Schematic showing the technique for cutting the crystal lattice with a Miller plane $(h k l)$ with direction $\v{N}$ passing through the point $\v{n} = c\v{N} = c (h  \v{b}_1 + k \v{b}_2 + l \v{b}_3$). The selected region comprises every point in the lattice except any point from the plane and above, with direction \v{N} (the region that was cut).}
        \label{select_points}
    \end{figure}
    is the set containing all lattice points lying ``below"  the plane $\Pi(c)$. See Figure~\ref{select_points} for a schematic representation of this procedure.  
    If now we want to cut the lattice by a Miller plane, we first need to expand out the Miller indices $(hkl)$ using the reciprocal vectors, so that the normal vector has the form $\v{N}_{(hkl)} = h \v{b}_1 + k \v{b}_2 + l \v{b}_3$. The cutting criterion will be the same as before in the set definition of Eq.~(\ref{cut_with_N}) except that now the normal vector we use will be $\v{N}_{(hkl)}$.     
    
    The plane periodicity $T$ along $\v{N}_{(hkl)}$ can be computed as:
    \begin{equation}
        T = \frac{2\pi}{\norm{\v{N}_{(hkl)}}} = \frac{2\pi}{\norm{h \v{b}_1 + k \v{b}_2 + l \v{b}_3}}\;.
    \end{equation}
    This quantity gives the distance between two adjacent Miller planes with the same indices along the direction $\v{N}_{(hkl)}$. A derivation of this formula is given in \ref{planeper}. 
    If, for example, we take $(h k l) = (1 0 0)$, then we have: 
    \begin{equation*}
        \eqalign{
            T =  \frac{2\pi}{h\norm{\v{b}_1}} &= 
            \frac{2\pi}{ 2 \pi
                \norm{
                    \frac{
                        \v{a}_2 \times \v{a}_3}{\v{a}_1 . (\v{a}_2 \times \v{a}_3)
                    } 
                }
            } \\
            & = \norm{\v{a}_1} \cos(\upsilon)\;,
        }
    \end{equation*}
    so that the periodicity in the direction normal to the $(100)$ face, i.e. the distance between two adjacent planes with Miller indices $(100)$, will then be $a \cos(\upsilon)$; where $\upsilon$ is the angle between $\v{a}_1$ and $\v{a}_2 \times \v{a}_3$ and $a$ is the lattice parameter in Eq.~(\ref{first_lattice_param}). 
    
    \begin{figure}
        \centering
        \begin{tikzpicture}
            [decoration={markings,mark=between positions 0 and 1 step 8pt with { \draw [fill,gray!80] (0,0) circle [radius=1pt];}}]  
            \path[postaction={decorate},xshift=-2.0cm]   (-1.5,1.5) to (1.5,-1.5);
            \path[postaction={decorate},xshift=-1cm]  (-1.5,1.5) to (1.5,-1.5); 
            \path[postaction={decorate},xshift=-0cm]  (-1.5,1.5) to (1.5,-1.5);
            \path[postaction={decorate},xshift=1.0cm]  (-1.5,1.5) to (1.5,-1.5);
            \path[postaction={decorate},xshift=2cm]  (-1.5,1.5) to (1.5,-1.5);
            
            \draw[line width=2pt,blue,|-|](0,0)--(0.5,0.5) node[anchor=south]{$T$};
            \draw[line width=1pt,blue,-stealth](-0.0,-0)--(2,2) node[anchor=south]{$\v{N}_{(hkl)}$};
            
        \end{tikzpicture}
        \caption{Scheme showing the plane periodicity $T$ (blue segment) taken as the minimun distance between to contiguous planes from a particular family of equivalent $(hkl)$ planes.}
        \label{plane_periodicity}
    \end{figure}
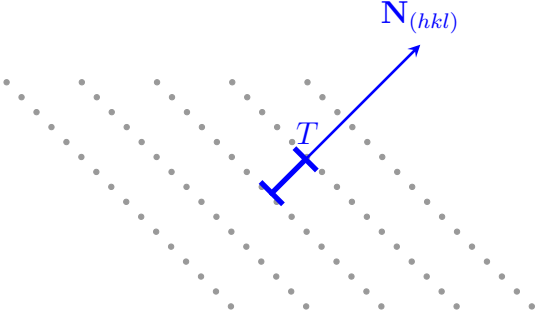

    We can hence normalize the $\v{N}_{(hkl)}$ direction and cut (select the points below the $(hkl)$ plane) using the expression in Eq.~(\ref{cut_with_N})
    with $\v{n} = t T \widehat{\v{N}}$ and $\widehat{\v{N}}$ the unit normal. As $t$ varies, this will select all planes along the $\v{N}_{(hkl)}$ direction towards the origin ``below" the $(hkl)$ plane. This constitutes a full procedure to cut a crystal shape using Miller planes. Note that the only input variables apart from the lattice parameters are the $(hkl)$ indices and the scaling factor $t$. If $\left\{\v{r}^{b}_{j}\right\}_{j = 1,N_b}$ are the coordinates of the crystal basis expressed in fractional coordinates of the lattice vectors, the full crystal structure system coordinates will be $\left\{\v{r}_{i} + x^{b}_{j} \v{a}_1 + y^{b}_{j} \v{a}_2 + z^{b}_{j} \v{a}_3\right\}_{1\leq i \leq N, 1 \leq j \leq N_b }$.
    
    \subsection{Computing PBC vectors}
    \label{extendedpbc}
    In order to construct a crystal slab, we need to define all the plane boundaries that will form the parallelepiped or PBC cell. To do so, we cut using a total of six Miller planes. 
    Given a Miller plane $(h k l)$ and its normal vector $\v{N}_{(hkl)}$, we seek to find two additional Miller planes $(h_1 k_1 l_1)$ and $(h_2 k_2 l_2)$, yielding two additional normal vectors $\v{N}_1$ and $\v{N}_2$, such that all three normal vectors, $\v{N}_{(hkl)}$, $\v{N}_1$ and $\v{N}_2$ are perpendicular to each other with respect to the reciprocal basis $\mathcal{B} \equiv \left\{\v{b}_1,\v{b}_2,\v{b}_3 \right\}$. Note that the vectors $\v{N}_{(hkl)}$, $\v{N}_1$ and $\v{N}_2$ when expressed in the canonical basis set $\mathcal{E}$ might not be orthogonal to one another since $\mathcal{B}$ is not necessarily an orthogonal basis set, e.g. in the case of a monoclinic or triclinic unit cell.
    If we have a Miller plane $(h k l)$ with normal vector $\v{N}_{(hkl)} =h \v{b}_1 + k \v{b}_2 + l \v{b}_3$, we can, without loss of generality, assume that $h \neq 0$ and pick a first vector $\v{N}_1 = h_1 \v{b}_1 + k_1 \v{b}_2 + l_1 \v{b}_3$ perpendicular to it in the basis $\mathcal{B}$ by setting  $ h_1 = (- k k_1 - l l_1)/h$ for given $k_1$ and $l_1$. This is an immediate consequence of solving for $h_1$ in the equation $(h_1,k_1,l_1) \cdot (h_2,k_2,l_2) = 0$.
    Since $k_1$ and $l_1$ are free parameters, we choose to set them to 1 and 0 respectively to get  $h_1 = -k/h$. We can also compute the entries of another vector $\v{N}_2 = (h_1 \v{b}_2 + k_2 \v{b}_2 + l_2 \v{b}_3)$ orthogonal to $\v{N}$ and $\v{N}_1$ that will have components: 
    \begin{equation}
        \label{eqn:N dot N_2}
        \eqalign{    
            h_2 &=  \frac{k^2 l/h^3}{k^2/h^2 + 1} - l/h \\
            k_2 &= \frac{-k l/h^2}{k^2/h^2 +1} \\
            l_2 &= 1 
        }
    \end{equation}
    Here $l_2$ is also a free parameter that was set to 1 for convenience. The equations in Eq.~(\ref{eqn:N dot N_2}) come from solving for $h_2$, $k_2$, $l_2$ using the system of equations given by $\left\{ [\v{N}_{(hkl)}]_{\mathcal{B}} \cdot [\v{N}_2]_{\mathcal{B}} = 0, \,  [\v{N}_1]_{\mathcal{B}}\cdot [\v{N}_2]_{\mathcal{B}} = 0\right\}$, where we use $[.]_{\mathcal{B}}$ to denote that a vector is expressed in its $\mathcal{B}$ basis representation.  In the case where $h$ is zero we can always permute two coordinates, apply the formulas and permute back. If fractional numbers are obtained from computing $h_1$, $h_2$, $k_2$ or $l_2$, one can always divide by the minimum value that was obtained for those entries that are non-zero.
    
    
    %
    A pseudocode implementing this procedure can be found in \ref{pseudoPerp}. There are many alternative ways of obtaining two orthogonal vectors to a particular $\v{N}_{(hkl)}$ direction; here we have only proposed one such technique. This way of constructing orthogonal directions allows us to define slabs that are bounded 
    by a parallelepiped constructed out of the corresponding Miller planes. For example, if $(h k l) = (1 1 0)$, then the procedure just described produces $(h_1 k_1 l_1) = (\overline{1} 1 0)$ and $(h_2 k_2 l_2) = (0 0 \overline{1})$. 
    
\begin{figure}
        \centering
        \begin{tikzpicture}
            [decoration={markings,mark=between positions 0 and 1 step 6pt with { \draw [fill,gray!80] (0,0) circle [radius=1pt];}}]  
            \path[postaction={decorate},xshift=-2.0cm]   (-1.5,1.5) to (1.5,-1.5);
            \path[postaction={decorate},xshift=-1.5cm]  (-1.5,1.5) to (1.5,-1.5);
            \path[postaction={decorate},xshift=-1.0cm]  (-1.5,1.5) to (1.5,-1.5);
            \path[postaction={decorate},xshift=-0.5cm]  (-1.5,1.5) to (1.5,-1.5);
            \path[postaction={decorate}]  (-1.5,1.5) to (1.5,-1.5);
            \path[postaction={decorate},xshift=0.5cm]  (-1.5,1.5) to (1.5,-1.5);
            \path[postaction={decorate},xshift=1.0cm]  (-1.5,1.5) to (1.5,-1.5);
            \path[postaction={decorate},xshift=1.5cm]  (-1.5,1.5) to (1.5,-1.5);
            \path[xshift=2.0cm,postaction={decorate, decoration={markings,mark=between positions 0 and 1 step 6pt with { \draw [fill,green] (0,0) circle [radius=1pt];}}}]  (-1.5,1.5) to (1.5,-1.5);   
            
            \draw[<->] (-2,0)--(3,0) node[right]{$x$};
            \draw[<->] (0,-2)--(0,2) node[above]{$y$};
            
            \draw[line width=2pt,blue,-stealth](0,0)--(1,1) node[anchor=south]{$\v{n}^+$};
            \draw[line width=2pt,blue,-stealth](1,1)--(2,2) node[anchor=south]{$\v{n}$};
            \draw[line width=2pt,red,-stealth](0,0)--(2,0) node[anchor=north east]{$\v{v}$};
            \draw[line width=2pt,red,-stealth](0,0)--(-2,0) node[anchor=north east]{$\v{v}$};
            \draw[line width=2pt,blue,-stealth](0,0)--(-1,-1) node[anchor=north east]{$\v{-n}$};
        \end{tikzpicture}
        \caption{Scheme showing the relationship between the vector normal to the crystal face, $\v{n}$, and a vector \v{v} in the reciprocal direction such that $\v{p} = 2\v{v}$, where $\v{p}$ is one of the PBC vectors.}
        \label{scheme}
    \end{figure}
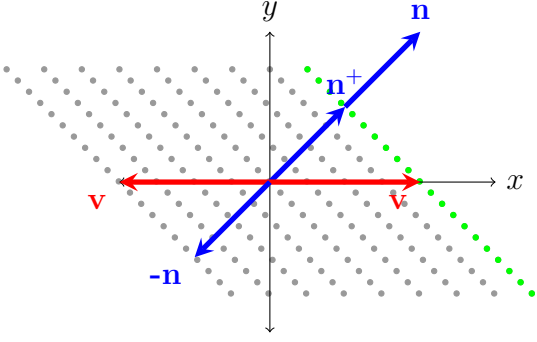

    We then cut a slab using 
    the procedure explained above which finds the appropriate bounding planes of a parallelpiped. The faces of this parallepiped will have normal vectors: 
    \begin{equation}
        \label{eq:slab-normals}
        \begin{array}{cccc}
            \v{n}^+ &= t T\widehat{\v{N}}, & \v{n}^{-} &= -t T \widehat{\v{N}} \\ 
            \v{n}_{1}^+ &= s T_{1} \widehat{\v{N}}_1, & \v{n}_{1}^- &= -s T_{1}\widehat{\v{N}}_1 \\
            \v{n}_{2}^+ &= r T_{2} \widehat{\v{N}}_2, & \v{n}_{2}^- &= -r T_{2} \widehat{\v{N}}_2 
        \end{array}
    \end{equation}

    \noindent and by varying over the parameters $t$, $s$ and $r$, we change the aspect ratio as well as the volume of the slab we construct. Once the slab is cut, we then need to find its PBC vectors knowing only the $h,k,l$ indices with which the faces of the solid were cut. Suppose we have a situation like the one depicted in Figure~\ref{scheme} where we have performed two cuts using $\v{n}^+$ and $\v{n}^{-}$. 
    The vector $\v{v}$ that makes up half of the PBC vector ends right on the surface of the solid; on the plane with normal vector $\v{n}$. The vector $\v{v}$ is then in the ``reciprocal'' direction and can be computed as: $\hat{\v{v}} = (\v{n}_1 \times \v{n}_2)/\norm{ \v{n}_1 \times \v{n}_2}$. Similarly,
    \begin{align*}
        \hat{\v{v}}_1 &= (\v{n} \times \v{n}_2)/\norm{ \v{n} \times \v{n}_2}\;,\\
        \hat{\v{v}}_2 &= (\v{n} \times \v{n}_1)/\norm{ \v{n} \times \v{n}_1}\;.
    \end{align*}
    Then, if $\v{v} = x  \hat{\v{v}}$, 
    we have that $(x  \hat{\v{v}} - \v{n}) \cdot \v{n} = 0$, from which $x$ can be solved, leading to $x = |\v{n}|^2/ |\hat{\v{v}}\cdot \v{n}|$. We therefore take the PBC vector $\v{p}$ to be, $\v{p} = 2( |\v{n}|^2 \hat{\v{v}})/ |\hat{\v{v}}\cdot \v{n}|$ and similarly:
    \begin{align*}
        \begin{split}
            \v{p}_1 =  2( |\v{n}_1|^2 \hat{\v{v}}_1)/ |\hat{\v{v}}_1\cdot \v{n}_1| \\
            \v{p}_2 =  2( |\v{n}_2|^2 \hat{\v{v}}_2)/ |\hat{\v{v}}_2\cdot \v{n}_2|
        \end{split} \;.
    \end{align*} 
    The result is the set of PBC vectors for our newly created slab.
    
    \subsection{Reorientation of the surface}
    We often would like to reorient a particular surface so that the surface normal vector aligns
    with a particular canonical vector. Let $\v{p}$ be the vector we would like to be aligned with $\v{e}_1 = (1,0,0)$.  This is important because one can easily and artificially include vacuum by enlarging
    the length of the $\v{p}$ vector yet keeping the same orientation. In order to do this reorientation we will need a linear transformation $\mathcal{M}$ that takes $\v{p} \rightarrow \; \norm{\v{p}}\v{e}_1$. One way to do this is to apply a transformation that gives us parameters $A$, $B$, $C$, $\Theta$, $\Phi$, $\Psi$ similar to what was explained for the lattice parameters in Section~\ref{background}. We define the transformation $\mathcal{T}^{>}: \left\{ \v{p},\v{p}_1,\v{p}_2 \right\} \rightarrow \left\{ A, B, C, \Theta, \Phi, \Psi \right\}$ by:
    \begin{equation}
        \eqalign{
            A &= \norm{\v{p}} \\
            B &= \norm{\v{p}_{1}} \\
            C &= \norm{\v{p}_{2}}
        }
    \end{equation}
    
    \begin{equation}
        \eqalign{
            \Theta &= \frac{360}{2 \pi} \arccos\left( (\v{p} \cdot \v{p}_1)/(A B) \right) \\
            \Phi &= \frac{360}{2 \pi} \arccos\left( (\v{p} \cdot \v{p}_2)/(A C) \right) \\
            \Psi &= \frac{360}{2 \pi} \arccos\left( (\v{p}_1 \cdot \v{p}_2)/(B C) \right) \;, \\
        }
    \end{equation}
    which is a coordinate system that is independent of the orientation of the slab. We can also define the ``back-transformation''  $\mathcal{T}^{<}:  \left\{ A, B, C, \Theta, \Phi, \Psi \right\} \rightarrow  \left\{ \v{p}^{||},\v{p}^{||}_1,\v{p}^{||}_2 \right\}$ that will give back a reoriented set of PBC vectors parallel to $\v{e}_1$ as follows:
    \begin{equation}
        \v{p}^{||} = A \v{e}_1 = A (1,0,0)
    \end{equation}
    
    \begin{equation}
        \v{p}^{||}_{1} = B (\cos(\Theta), \sin(\Theta), 0)
    \end{equation}
    
    \begin{equation}
        \eqalign{
            \v{p}^{||}_{2x} &=C \cos(\Phi) \\
            \v{p}^{||}_{2y} &=C \frac{\left( \cos(\Psi)-\cos(\Theta) \cos(\Phi) \right)} {\sin(\Theta)} \\
            \v{p}^{||}_{2z} &=\sqrt{\left(C^2 - (\v{p}^{||}_{2x})^2 - (\v{p}^{||}_{2y})^2 \right)}
        }
    \end{equation}
    
    Note that the composition of these two transformations  $\mathcal{M} = \mathcal{T}^{<}\mathcal{T}^{>}:  \left\{ \v{p},\v{p}_1,\v{p}_2 \right\} \rightarrow  \left\{ \v{p}^{||},\v{p}^{||}_1,\v{p}^{||}_2 \right\}$ 
    will lead to a vector $\v{p}^{||}$ aligned with $\v{e}_1$.
    The action of the transformation $\mathcal{M}$ can then be expressed using matrix algebra through:
    
    \begin{align}
        \begin{split}
            \mathcal{M} \times & \begin{pmatrix} 
                \v{p}_{x} & \v{p}_{1x} & \v{p}_{2x} \\ 
                \v{p}_{y} & \v{p}_{1y} & \v{p}_{2y} \\
                \v{p}_{z} & \v{p}_{1z} & \v{p}_{2z} 
            \end{pmatrix} \\
            & \hspace{1cm} =
            \left( \begin{matrix} 
                \v{p}^{||}_{x} & \v{p}^{||}_{1x} & \v{p}^{||}_{2x} \\ 
                \v{p}^{||}_{y} & \v{p}^{||}_{1y} & \v{p}^{||}_{2y} \\
                \v{p}^{||}_{z} & \v{p}^{||}_{1z} & \v{p}^{||}_{2z} 
            \end{matrix} \right) \;,
        \end{split}
    \end{align}
    so that,
    \begin{align}    
        \begin{split}
            \mathcal{M} &= \left( \begin{matrix} 
                \v{p}^{||}_{x} & \v{p}^{||}_{1x} & \v{p}^{||}_{2x} \\ 
                \v{p}^{||}_{y} & \v{p}^{||}_{1y} & \v{p}^{||}_{2y} \\
                \v{p}^{||}_{z} & \v{p}^{||}_{1z} & \v{p}^{||}_{2z} 
            \end{matrix} \right) \\
            & \hspace{1cm} \times \left( \begin{matrix} 
                \v{p}_{x} & \v{p}_{1x} & \v{p}_{2x} \\ 
                \v{p}_{y} & \v{p}_{1y} & \v{p}_{2y} \\
                \v{p}_{z} & \v{p}_{1z} & \v{p}_{2z} 
            \end{matrix} \right)^{-1}  \;.
        \end{split}
    \end{align}
    
    Therefore the coordinates of every point $\v{r}$ from the original slab can be transformed using $\mathcal{M}$ as  $(\v{r}^{||})^t = \mathcal{M} \v{r}^t$. 
    
    \section{Building non-periodic solids}
    \label{nonper}
    The technique to cut by planes explained above allows us to construct any crystalline convex polyhedron just by simply using a list of planes and their distance to the origin. These planes could have any direction $\v{N}$ (or if they are Miller planes, any specific $\v{N}_{(hkl)}$ direction). A crystal slab could also be viewed as a particular crystalline convex polyhedron in which 
    the shape is a parallelepiped. 
    
    Crystalline convex polyhedra can also be cut out by taking some aspect of crystal growth into consideration in order to predict and visualize equilibrium crystal shapes. 
    Observations by Wulff concluded that there is a relationship between the extension of the exposed crystal surface and the speed at which it grows \cite{Wulff1901-mn,Mutter_undated-kz}. Different surfaces grow at different rates which ultimately determines the extension of each of the exposed surfaces. Wulff realized that the slower a surface grows, the more extended it will appear in the final crystal shape when thermodynamic equilibrium is reached \cite{Wulff1901-mn,Mutter_undated-kz}. 
    
\begin{figure*}
        \centering
        \begin{tikzpicture}
            \coordinate (O) at (0, 0);
            \node[inner sep=0pt] (benzene) at (O)
            {\includegraphics[width=0.9\textwidth]{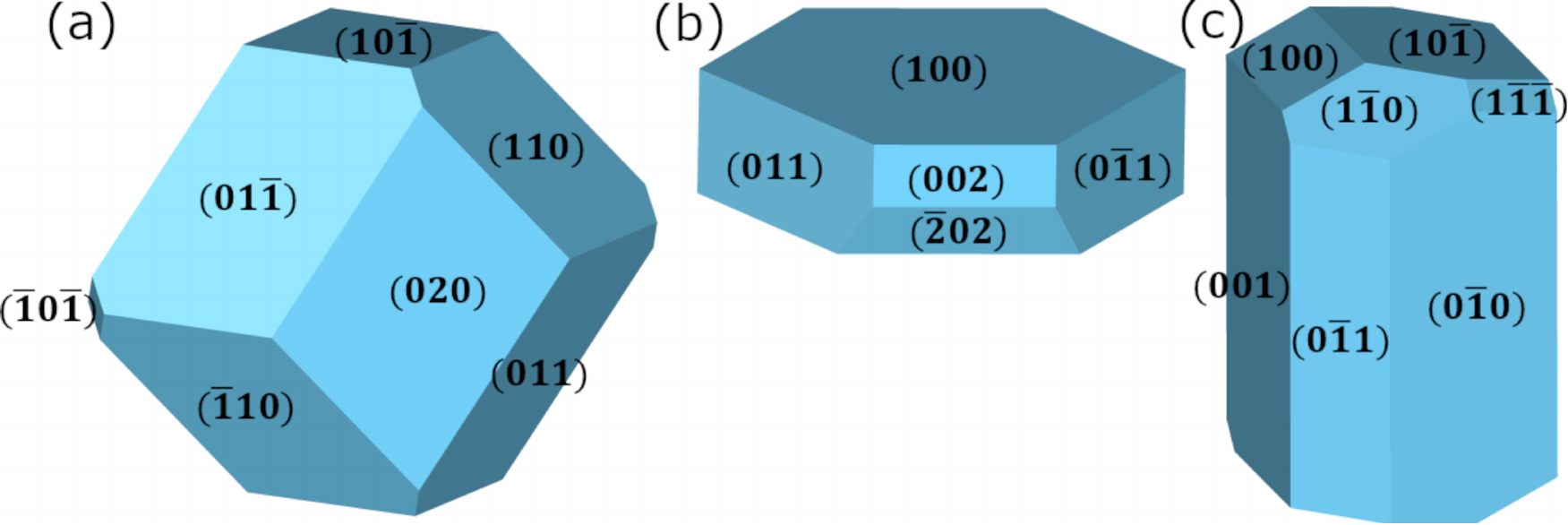}};
            
        \end{tikzpicture}
        \caption{Wulff structures for (a) monoclinic $\beta$-HMX, (b) monoclinic PFBA, and (c) triclinic IDOX based on the BFDH theory. The figure indicates the indices of only the front faces. The back faces are symmetrically opposite to the front faces and have the same indices in the opposite direction.}
        \label{wulff2}
    \end{figure*}

    If a face $a$ with direction $\v{N}_a$ grows faster than another face $b$ with direction $\v{N}_b$ then, to have an idea of the crystal shape, the plane boundary with normal $\v{N}_a$ will be placed much farther away from the origin than the plane boundary with $\v{N}_b$. In both cases the distances to the origin at which a plane should be used to cut the crystal will be directly proportional to the growth rate. Donnay and Harker's law \cite{Donnay1937-au} expresses that the 
    surface growth rates are inversely proportional to the distance between adjacent $(hkl)$ planes $d_{hkl}$. This law, combined with Wulff's observations leads to the Bravais-Friedel-Donnay-Harker (BFDH) criterion  \cite{Docherty1991-ne} which gives a rule for constructing the equilibrium crystal shape based solely on crystallographic data where plane positions are set such that the distance to the origin is proportional to $1/d_{hkl}$.
    The Wulff crystal convex polyedron based on BFDH method could be hence defined as: 
    
    \begin{equation}
        \{\v{r} \in \mathbb{R}^3 | (\v{r} - \frac{\alpha}{d_{hkl}} \hat{\v{N}})\cdot \hat{\v{N}}  \,\, , \forall \,\, (hkl) \,\, \mathrm{planes} \} 
    \end{equation}
    where:
    \begin{equation}
        \hat{\v{N}}  =  \frac{h \v{b}_1 + k \v{b}_2  + l \v{b}_3}{\norm{ h \v{b}_1 + k \v{b}_2  + l \v{b}_3}}
    \end{equation}
    and $\alpha$ is some  proportionality constant. Using our code is thus trivial to build Wulff structures predicted by the BFDH theory (including plane selection rules based on reflection conditions\cite{cryst_tables}). In Figure~\ref{wulff2}, we illustrate this by comparing the predicted morphologies of an energetic crystal, monoclinic $\beta$-1,3,5,7-tetranitro-1,3,5,7-tetrazoctane ($\beta$-HMX, space group $P2_1/n$, CSD~\cite{CSD} entry OCHTET13), and two proposed surrogates~\cite{yeager2018development}, triclinic 5-Iodo-2'-deoxyuridine (IDOX, $P1$, IDOXUR) and monoclinic 2,3,4,5,6-Pentafluorobenzamide (PFBA, $P2_1/c$, VATNOU). We note that the Wulff structure for $\beta$-HMX shown in Figure~\ref{wulff2} is the same as presented in Refs.~\cite{Gallagher2017-je,Gallagher2014-vq}, which is also based on the BFDH theory. The three crystals exhibit markedly different shapes and exposed faces, due to distinct crystal structures and symmetries.  
    
    More complex models, incorporating for instance explicitly calculated surface energies, can also be implemented.
    The obtained shapes determine the most important crystalline faces that can then be compared to experimental observations of real crystals~\cite{Gallagher2017-je}.

    \section{Building a crystal slab with orthogonal PBC vectors}
    \label{ortho}
    In this section we detail a new technique to generate an orthogonal slab exposing a desired $(hkl)$ ``front'' plane while maintaining periodicity of the lattice. Here we have the opposite viewpoint of the previous section, we now want to first construct the PBC vectors, and then from these, construct the orthogonal slab. The periodicity is maintained up to a user-specified tolerance of how well the orthogonal slab conforms to the lattice vectors. How this can be measured will be explained below. All vectors in this section are expressed in the canonical basis.

    To construct the slab, three PBC vectors must be selected. The first vector, $\v{p}$, will have the direction of the prescribed normal vector $\v{N}_{(hkl)}$ of the given $(hkl)$ plane so that $\v{p} = c \v{N}_{(hkl)}$ for some scalar $c$. The two remaining PBC vectors, $\v{p}_1$ and $\v{p}_2$, are chosen so that all three PBC vectors are orthogonal to one another. 
    To find these orthogonal vectors we use a similar procedure to the one described in Section~\ref{extendedpbc}. Having the vector $\v{p}$ in hand, we choose the second vector, $\v{p}_1$, so that $\v{p} \cdot \v{p}_1 = 0$. For this we define the $x$-component of $\v{p}_1$ as 
    \begin{equation}
        \begin{aligned}
            \v{p}_{1,x} &= -\frac{\v{p}_{y}}{\v{p}_{x}} \;,            
        \end{aligned} \label{eq:n2}
    \end{equation}
    where the free parameters $\v{p}_{1,y}$, and $\v{p}_{1,z}$ have been set to $1$ and $0$ respectively for convenience. Note that if $\v{p}$ was chosen such that $\v{p}_{x} = 0$, a relabeling of the axes can be made. The third vector $\v{p}_2$ can be generated by requiring that $\v{p} \cdot  \v{p}_2 = 0$ and $\v{p}_1\cdot \v{p}_2 = 0$ to get:
    \begin{equation}
        \begin{aligned}
            %
            %
            %
            %
            \v{p}_{2,x} &= \frac{\v{p}_{y}^2 \v{p}_{z}/\v{p}_{x}^3}{\v{p}_{y}^2/\v{p}_{x}^2 +1 } - \frac{\v{p}_{z}}{\v{p}_{x}}\, \\
            \v{p}_{2,y} &= -\frac{\v{p}_{y}/\v{p}_{x}^3}{\v{p}_{y}^2/\v{p}_{x}^2 +1 }\;, 
        \end{aligned}
    \end{equation}
    where the free parameter $\v{p}_{2,z}$ has been set to 1. 
    Note that by setting different parameters one will obtain different equations which will give orthogonal vectors with different directions. Once a pair of orthogonal vectors $\bf{p}_1$ and $\bf{p}_2$ are obtained, a rotation around the vector $\bf{p}$ can be applied to explore other directions. 
    \subsection{Minimum translation, integer search}\label{ortho-integer-search}
    Given the vectors $\v{p}$, $\v{p}_1$, $\v{p}_2$ just constructed, we now seek to express a certain rescaling of each of the vectors as a linear combination of the lattice vectors in order to comply with periodicity. For any real number $t>0$, a vector $t \v{p}$ fully respects the periodicity of the unit cell if $t\v{p}$ can be written as an integer linear combination of the lattice vectors. This condition can be represented by the linear system of equations:
    \begin{align}
        \begin{split} \label{eq:orth2lat_linearsystem}
            t \v{p}_x &=  {\v{a}_{1,x}} a + {\v{a}_{2,x}} b + {\v{a}_{3,x}} c\\
            t \v{p}_y &=  {\v{a}_{1,y}} a + {\v{a}_{2,y}} b + {\v{a}_{3,y}} c\\
            t \v{p}_z &=  {\v{a}_{1,z}} a + {\v{a}_{2,z}} b + {\v{a}_{3,z}} c
        \end{split} \;,
    \end{align}
    where $a$, $b$, and $c$ are the unknown scaling, or translation,  coefficients. We therefore should choose the parameter $t$ such that $a$, $b$, and $c$ are as close as to integers as possible. For a specific choice of $t$, we define
    \begin{align} \label{eq:orth2lat}
        \v{M} &= 
        \begin{pmatrix}
            \v{a}_{1,x} & \v{a}_{2,x} & \v{a}_{3,x} \\
            \v{a}_{1,y} & \v{a}_{2,y} & \v{a}_{3,y} \\
            \v{a}_{1,z} & \v{a}_{2,z} & \v{a}_{3,z}
        \end{pmatrix} \;,
    \end{align}
    to obtain the matrix equation 
    \begin{align} \label{eq:orth2lat_matrix}
        \begin{pmatrix} a(t) \\ b(t) \\ c(t) \end{pmatrix} = t \v{M}^{-1} \v{p} \;,
    \end{align}
    which solves Eq.~(\ref{eq:orth2lat_linearsystem}). We can then determine for each parameter $t$ in some pre-defined range how close the $a(t)$, $b(t)$, and $c(t)$ translation coefficients are to integers. In other words, the optimal choice of $t$ needs to result in translation coefficients $a(t)$, $b(t)$ and $c(t)$ that minimize the quantity:
    \begin{align}
        \begin{split} \label{error}
            \varepsilon(t) &= |a(t)-\text{Rnd}(a(t))| + |b(t)-\text{Rnd}(b(t))| \\
            & \hspace{2cm} + |c(t)-\text{Rnd}(c(t))| \;,
        \end{split}
    \end{align}
    where the round-off function 
    \begin{align}
        \mathrm{Rnd}(x) = \begin{cases} 
            \lfloor x \rfloor &, x-\lfloor x \rfloor < \tfrac{1}{2}\\
            \lceil x \rceil &, x-\lfloor x \rfloor \ge \tfrac{1}{2}
        \end{cases}
    \end{align}
    takes the closest integer to the argument $x$. Here, $\lfloor x \rfloor$ and $\lceil x \rceil$, denote the floor and ceiling of the real number $x$, respectively. The function $\varepsilon$ has a maximum of 1.5 at non-integers and a minimum of 0 when $a,b$ and $c$ are all integers, indicating a perfect periodic arrangement in the given direction. 
    
    It is in fact the case that $t$ in Eq.~(\ref{eq:orth2lat_matrix}) can be chosen such that $a$, $b$ and $c$ are exactly integers, however this choice of $t$ may be unacceptably large. Assuming that the entries of $\v{M}$ and $\v{p}$ are rational, the determinant formula for a $3 \times 3$ matrix says that $\v{M}^{-1}$ should also have rational entries. Hence, if $(\v{M}^{-1})_{ij} = m_{ij}/q_{ij}$ and $\v{p}_i = u_i/v_i$ are such that the numerators and denominators are relatively prime, i.e. have no common factors, then selecting 
    $t = \prod_{\{ij|m_{ij}\neq0 \}} q_{ij} \prod_{\{i|u_i\neq0\}} v_i$
    will ensure that the $a,b,c$ translation coefficients are all integers. Our approach will thus need to balance the size of the slab with the error in the periodicity.
    
    In order to construct the desired orthogonal slab, the above procedure is carried out for each PBC vector over the $t$, $s$ and $r$ parameter space such that $\varepsilon$ is minimized. Using Eq.~(\ref{eq:orth2lat_matrix}), we do a complete search over the parameter space for some specified range of values.
    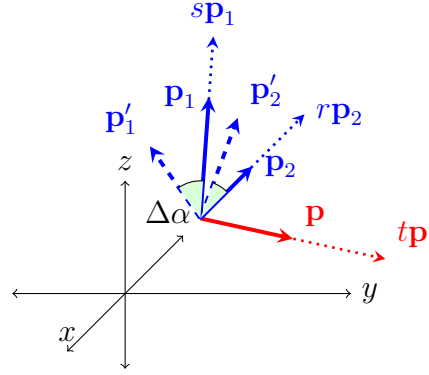
\begin{figure}
        \centering
        \begin{tikzpicture}
            
            \draw[<->] (0,0,-2)--(0,0,2) node[above]{$x$};
            \draw[<->] (-1.5,0,0)--(3,0,0) node[right]{$y$};
            \draw[<->] (0,-1,0)--(0,1.5,0) node[above]{$z$};

            \coordinate (a) at (1.5,3,1);
            \coordinate (b) at (1,1,0);
            \coordinate (c) at (0.96,2.59,1.64);
            
            \coordinate (d) at (1.5,1.5,-0.5);
            \coordinate (e) at (1,1,0);
            \coordinate (f) at (1.33,2.17, -0.435);
            
            \draw[line width=1.5pt,red,-stealth](1,1,0)--(3,1.5,2) node[anchor=south west]{$\v{p}$};
            \draw[line width=1.0pt,red,-stealth, dotted](1,1,0)--(5,2,4) node[anchor=south west]{$t\v{p}$};
            \draw[line width=1.5pt,blue,-stealth](1,1,0)--(1.5,3,1) node[anchor=east]{$\v{p}_1$};
            \draw[line width=1pt,blue,-stealth,dotted](1,1,0)--(1.75,4,1.5) node[anchor=south]{$s\v{p}_1$};
            \draw[line width=1.5pt,blue,-stealth](1,1,0)--(1.5,1.5,-0.5) node[anchor=west]{$\v{p}_2$};
            \draw[line width=1pt,blue,-stealth,dotted](1,1,0)--(2,2,-1) node[anchor=west]{$r\v{p}_2$};
            \draw[line width=1.5pt,blue,-stealth,dashed](1,1,0)--(0.96,2.59,1.64) node[anchor=south east]{$\v{p}_1'$};
            \draw[line width=1.5pt,blue,-stealth,dashed](1,1,0)--(1.33,2.17, -0.435) node[anchor=south west]{$\v{p}_2'$};
            \draw pic[draw,fill=green!15,angle radius=0.5cm,"$\Delta \alpha$" shift={(-3.5mm,-2mm)}] {angle=a--b--c};
            \draw pic[draw,fill=green!15,angle radius=0.5cm] {angle=d--e--f};
        \end{tikzpicture}
        \caption{Scaled PBC vectors $t\v{p}$, $s\v{p}_1$ and $r\v{p}_2$ obtained from a parameter search are shown. In order to find potentially better $s$ and $r$ scaling coefficients, a rotation is done around the vector $\v{p}$ by angle $\Delta \alpha$ of $\v{p}_1$ and $\v{p}_2$ to $\v{p}_1'$ and $\v{p}_2'$. }
        \label{rotate around p}
    \end{figure}
    Both the $\v{p}_1$ and $\v{p}_2$ vectors can also be rotated by an angle $\Delta \alpha$ about $\v{p}$ to further improve the error during the search. This may result in potentially better $s$ and $r$ parameters, where better here means, lower values of $s$ and $r$ for a given tolerance since that would require a smaller number of atoms needing to be simulated (see Figure~\ref{rotate around p}). Equation \ref{error} is therefore calculated for each angle $\alpha$, at every value of $s$ and $r$. The total dimension of the parameter space to be searched is then four --- $t$, $s$, $r$ and $\alpha$ --- and the time-to-solution will scale as $N_t N_s N_r N_{\alpha} $, where $N_{\alpha} = 180/(\Delta \alpha)$ and 
    $N_{x} = (x_{\rm max} - x_{\rm min})/\Delta x$ for $x \in \{t,s,r\}$. The max and min values for each of the parameter search can be systematically increased or decreased if the error falls above the desired tolerance after a complete search. 
    
    Once the PBC vectors are rescaled by the $t$, $s$ and $r$ values that were found, we then cut the slab by determining which lattice points lie inside the parallepiped defined by the orthogonal PBC vectors. 
    The idea is to first transform the lattice point coordinates into their PBC vector basis representation. Given a point $\v{r}$ expressed in the canonical basis set, we want a transformation $T$ such that $T \v{r} = r^{\rm PBC}_x  \v{p} + r^{\rm PBC}_y \v{p}_1 + r^{\rm PBC}_z \v{p}_2$, that is, we want to solve the system
    \begin{align}
        \begin{split} \label{cut}
            r_{x} &= r^{\rm PBC}_x \v{p}_{x} + r^{\rm PBC}_y\v{p}_{1x} + r^{\rm PBC}_z\v{p}_{2x} \\
            r_{y} &= r^{\rm PBC}_x \v{p}_{y} + r^{\rm PBC}_y\v{p}_{1y} + r^{\rm PBC}_z\v{p}_{2y} \\
            r_{z} &= r^{\rm PBC}_x \v{p}_{z} + r^{\rm PBC}_y\v{p}_{1z} + r^{\rm PBC}_z\v{p}_{2z} 
        \end{split}\;.
    \end{align} 
    Therefore, $T$ will be given by the change of basis matrix
    \begin{equation}
        \label{cut1}
        T = \left( \begin{matrix} 
            \v{p}_{x} & \v{p}_{1x} & \v{p}_{2x} \\ 
            \v{p}_{y} & \v{p}_{1y} & \v{p}_{2y} \\
            \v{p}_{z} & \v{p}_{1z} & \v{p}_{2z} 
        \end{matrix} \right)  \;,
    \end{equation}
    so that by inverting $T$ and applying it to $\v{r}$, we can determine the PBC basis coordinates that we need. We then only accept the coordinates $(r^{\rm PBC}_x,r^{\rm PBC}_y,r^{\rm PBC}_z)$ if $0 < r^{\rm PBC}_\alpha < 1$, $\alpha = x,y,z$. Hence, the set of lattice points to be included in the slab are
    \begin{equation}\label{cut2}
        \left\{\v{r} : 0 < (T\v{r})_{x,y,z} < 1 \right\} \;.
    \end{equation}
    
    \subsection{Example of orthogonal PBC slabs}
    \label{examplesorth}
\begin{figure}
        \centering
        \includegraphics[width=1.1\columnwidth]{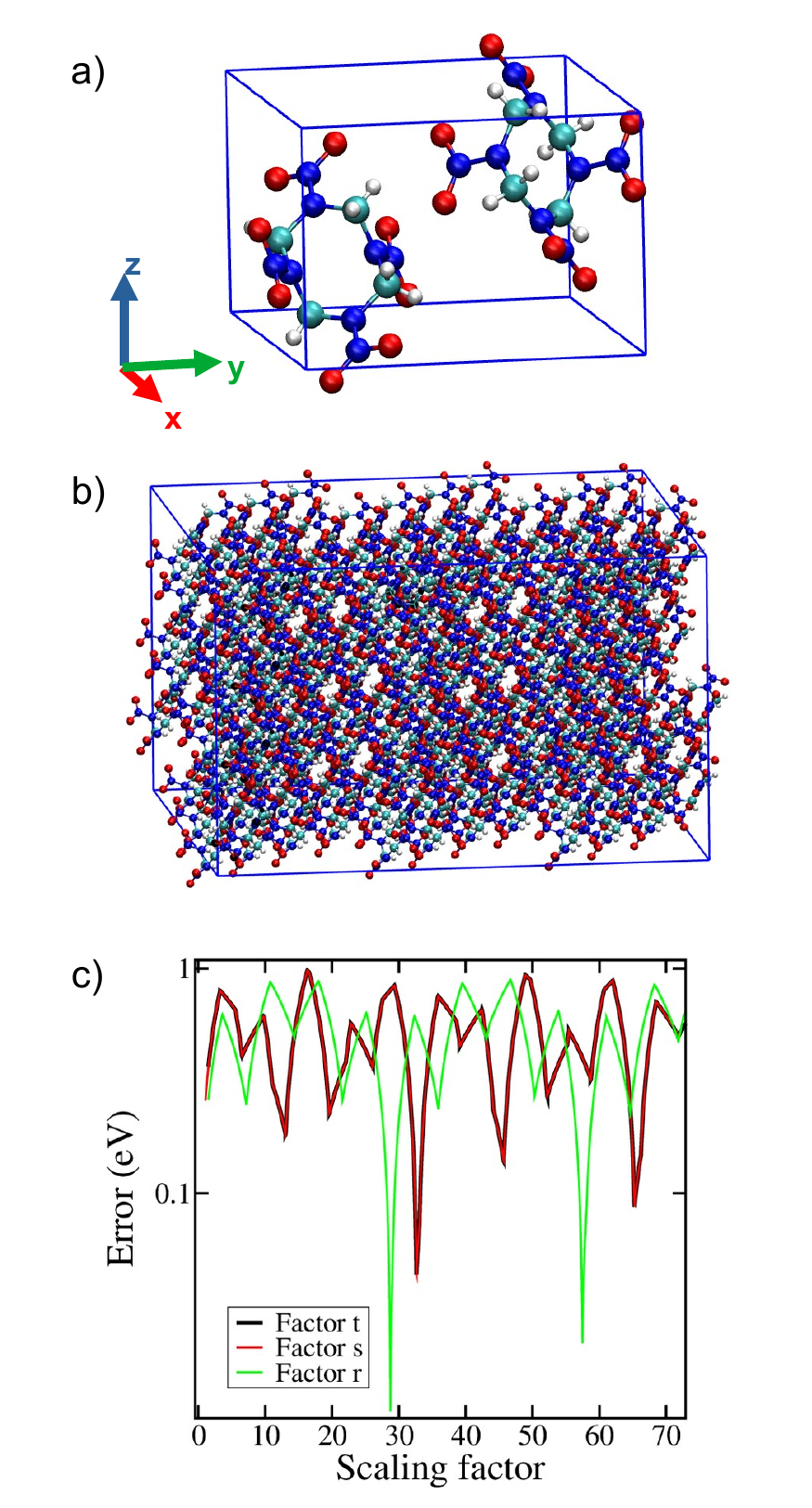}
        \caption{a) Schematic representation of the $\beta$-HMX  unit cell with its PBC box.  H, C, O, and N are represented with white, cyan, red, and blue spheres respectively. b). The resultant orthogonal PBC slab with minimal PBC error containing 118 $\beta$-HMX unit cells. The slab was reoriented such that the $x$-axis is parallel to the $(1,1,0)$ direction. c) The error function $\varepsilon$ as a function of the vector scaling factor. The black curve and red curve are on top of each other. No rotations are used to generate this plot.}
        \label{fig:HMX}
    \end{figure}

    In this section we demonstrate how to use the method described in Section~\ref{ortho-integer-search} to construct a slab with orthogonal PBC vectors by applying it to the $\beta$-HMX polymorph. $\beta$-HMX is a monoclinic crystal with a $P2_1/n$ space group. Figure \ref{fig:HMX}a depicts the $\beta$-HMX unit cell containing two HMX molecules for a total of 56 atoms. The lattice vectors expressed in the canonical basis set are:
    $\v{a}_1 = (6.53, 0, 0) $, $\v{a}_2 = (0, 11.02, 0)$, and $\v{a}_3 = (-1.61, 0, 7.18)$. %
    
    Given an initial direction of a desired exposed surface (not necessarily generated by a Miller plane) such as for example  $\v{p} = (1,1,0)$, the orthogonal normal vectors generated by the procedure explained in Section~\ref{ortho-integer-search}, are $\v{p}_1 = (-1, 1, 0)$ and $\v{p}_2 = (0, 0, 1)$. Applying a full parameter search with an error computed as in Eq.~(\ref{eq:orth2lat_matrix}) and Eq.~(\ref{error}), we find scaling factors of  $t=32.66$, $s = 32.66$, and $r=28.71$ for $\v{p}$, $\v{p}_1$, and $\v{p}_2$, respectively. These scalar lengths are found within a range of 1 to 100 using a step size of $\Delta = 10^{-3}$ and an error tolerance of $\delta = 10^{-1}$. Typically, the smaller we set the tolerance $\delta$, the larger the resulting volume will be.  Although this tolerance may not seem so small, we again must decide how large of a periodic cell we are willing to use. Substantially increasing the range of the parameters would result in smaller errors though at the cost of more atoms to simulate. In order to speed up the search for the scaling factor, one can use multiples of $\Delta = \text{min}(\frac{1}{a(t=1)} ,\frac{1}{b(t=1)}, \frac{1}{c(t=1)})$ so that $t = i \Delta$ where $i$ is an integer. This will ensure taking a ``large enough" step size. 
    
    In Figure~\ref{fig:HMX}c, we can see that the minimum error for both $\v{p}$ and $\v{p}_1$ is achieved at parameter value $t=s=32.6$, and the first minimum for $\v{p}_2$ is found at $r=7.17$. These minima are repeated at integer multiples of these values thereafter. For vector $\v{p}_2$, the error minima correspond to vector lengths that are approximately equal to multiples of the $z$ component value of the third lattice vector for HMX. Due to the error tolerance that was set, the minimum error value we select occurs at $r=28.71$.  Once the rescaling factors are obtained, a slab can be constructed using the procedure explained in Eqs.~\ref{cut} to \ref{cut2} using $\v{p} = 32.66 (1,1,0)$, $\v{p}_1 = 32.66  (-1, 1, 0)$, and $\v{p}_2 = 28.71  (0, 0, 1)$. The resulting slab is shown in Figure \ref{fig:HMX}b.

    \section{Conclusions}
    
    We have introduced a step-by-step procedure to cut a crystal slab from a crystal lattice which is typically needed for the study of both physical and chemical surface properties in molecular dynamics simulations. Our method can be used to recover the PBC vectors for a given crystal slab and generate any crystal-based convex polyedron such as Wulff structures. Additionally, we developed a procedure for directly constructing a crystal slab with orthogonal PBC vectors given a desired exposed surface. All these methods can be easily implemented as a preprocessing step for any computational chemistry code.  A first version of an in-house code \texttt{Los Alamos Crystal Cut} (LCC) that implements these methods can be found at \url{https://github.com/lanl/LCC}.

    \begin{appendix}

        \section{Pseudo-code for finding perpendicular directions to $(h k l)$}
        \label{pseudoPerp}
        The following pseudo-code takes three integers corresponding to $h$, $k$, and $l$ respectively and returns two orthogonal Miller indices $(h_1,k_1,l_1)$ and $(h_2,k_2,l_2)$ respectively. The function \textbf{permute}() is used to permute the indices in case $h$ = 0. The \textbf{permuteBack}() function gives back the correct order of the final indices. Finally, \textbf{minValNonZero}() takes the minimum non-zero value given the obtained indices. 
        
        \begin{algorithm}[H]
            \algrenewcommand\algorithmicfunction{\textbf{def}}
            \algrenewcommand\algorithmicend{.}
            \begin{algorithmic}
                \label{perphkl}
                \parskip 0.05cm
                {\fontsize{0.3cm}{0.3em}\selectfont 
                    \Function{get\_perp\_hkl}{$h,k,l$}:
                    \State \textbf{if}($h = 0$): \State\quad ($h',k',l'$) $\leftarrow$ \textbf{permute}($h,k,l$)
                    \State \textbf{else:}
                    \State\quad ($h',k',l'$) = ($h,k,l$)
                    \State  $k'_1 = 1$
                    \State  $l'_1 = 0$
                    \State  $h'_1 = -k'/h'$
                    \State  $h'_2 =  \frac{k'^2l'/h'^3}{k'^2/h'^2 + 1} - l'/h' $
                    \State  $k'_2 = \frac{-k'l'/h'^2}{k'^2/h'^2 +1} $
                    \State $l'_2  = 1$    
                    
                    \State \textbf{if}($h = 0$):        %
                    \State  \quad  \quad ($h,k,l$) $\leftarrow$ \textbf{permuteBack}($h',k',l'$)
                    \State \quad  \quad ($h_1,k_1,l_1$) $\leftarrow$ \textbf{permuteBack}($h'_1,k'_1,l'_1$)
                    \State \quad  \quad ($h_2,k_2,l_2$) $\leftarrow$ \textbf{permuteBack}($h'_2,k'_2,l'_2$)
                    \State \textbf{else}:
                    \State  \quad  \quad ($h,k,l$) = ($h',k',l'$)
                    \State \quad  \quad ($h_1,k_1,l_1$) = ($h'_1,k'_1,l'_1$)
                    \State \quad  \quad ($h_2,k_2,l_2$) = ($h'_2,k'_2,l'_2$)
                    
                    \State $m =$ \textbf{MinValNonzero}($h_1,k_1,l_1,h_2,k_2,l_2$)
                    \State ($h_1,k_1,l_1$) $\leftarrow$ ($h_1/m,k_1/m,l_1/m$)
                    \State ($h_2,k_2,l_2$) $\leftarrow$ ($h_2/m,k_2/m,l_2/m$)        
                    
                    \State    \textbf{return} $(h_1,k_1,l_1),(h_2,k_2,l_2)$                        
                    \EndFunction
                }       
            \end{algorithmic}
        \end{algorithm}
        
        \section{Wulff construction table of distances}
        The following table is a list of the planes and distances used to construct the shapes shown in Figure~\ref{wulff2}.
        
        \begin{table*}[htbp]
            \caption{Data used to build the Wulff structures shown in Figure~\ref{wulff2}. For each compound, we show planes of interest and the corresponding interplanar distance $d_{hkl}$ in \AA. For the highest indices, we limited ourselves to $h$=$k$=$l$=1, except when plane reflection rules dictated the use of $h$, $k$, and/or $l$=2 (see Ref.~\cite{cryst_tables}). Opposite planes with respect to the origin -- i.e. ($\bar{1}00$) for ($100$) -- are omitted for simplicity, and planes that are observed in the Wulff structure built following the BFDH theory are shown in bold font.}
            \begin{tabular}{|c|c|c|c|c|c|}
                \hline
                \multicolumn{2}{|l|}{$\beta$-HMX ($P2_1/n$)} & \multicolumn{2}{l|}{PFBA ($P2_1/c$)} & \multicolumn{2}{l|}{IDOX ($P1$)} \\ \hline
                Plane & $d_{hkl}$ & Plane & $d_{hkl}$ & Plane & $d_{hkl}$ \\ \hline
                $(002)$ & 3.59 & $\bf{(002)}$ & 4.72 & $\bf{(001)}$ & 8.73 \\ \hline
                $\bf{(020)}$ & 5.51 & $(020)$ & 3.10 & $\bf{(010)}$ & 6.51 \\ \hline
                $(200)$ & 3.19 & $\bf{(100)}$ & 12.53 & $\bf{(100)}$ & 4.58 \\ \hline
                $\bf{(011)}$ & 6.02 & $\bf{(011)}$ & 5.18 & $(011)$ & 4.66 \\ \hline
                $\bf{(01\bar{1})}$ & 6.02 & $\bf{(01\bar{1})}$ & 5.18 & $\bf{(01\bar{1})}$ & 6.05 \\ \hline
                $\bf{(101)}$ & 4.32 & $(202)$ & 3.49 & $(101)$ & 3.55 \\ \hline
                $\bf{(10\bar{1})}$ & 5.39 & $\bf{(20\bar{2})}$ & 4.13 & $\bf{(10\bar{1})}$ & 4.86 \\ \hline
                $\bf{(110)}$ & 5.52 & $(220)$ & 2.78 & $(110)$ & 3.40 \\ \hline
                $\bf{(1\bar{1}0)}$ & 5.52 & $(2\bar{2}0)$ & 2.78 & $\bf{(1\bar{1}0)}$ & 4.23 \\ \hline
                $(222)$ & 2.01 & $(111)$ & 4.63 & $(111)$ & 2.80 \\ \hline
                $(22\bar{2})$ & 2.42 & $(11\bar{1})$ & 4.96 & $(11\bar{1})$ & 3.73 \\ \hline
                $(2\bar{2}2)$ & 2.01 & $(1\bar{1}1)$ & 4.63 & $(1\bar{1}1)$ & 3.57 \\ \hline
                $(\bar{2}22)$ & 2.42 & $(\bar{1}11)$ & 4.96 & $\bf{(\bar{1}11)}$ & 4.08 \\ \hline
            \end{tabular}
            \label{tableWulff}
        \end{table*}
        
        \section{Plane periodicity}
        \label{planeper}
        In this section, we provide a formal argument that the plane periodicity (that is, the distance between two adjacent planes) can be computed using the formula:
        \begin{equation}
            T = d_{hkl} = \frac{2\pi}{\norm{h \v{b}_1 + k \v{b}_2 + l \v{b}_3}} \;.
        \end{equation}
        Let $\{.. \Pi_0 ... \Pi_i ... \Pi_N...\}$ be the collection of parallel planes with  Miller indices $(hkl)$. Within this set planes  $\Pi_i$ and  $\Pi_{i+1}$ are contiguous or neighboring planes. We find $T$ such that, $\mathrm{min} \, d(\Pi_i,\Pi_{i+1}) = T$, where $\Pi_i$ and $\Pi_{i+1}$ are two adjacent planes. Given the definition of Miller indices, without loss of generality we can say that both points $i\frac{\v{a}_1}{h}$ and $(i+1)\frac{\v{a}_1}{h} $ belong to $\Pi_i$ and $\Pi_{i+1}$ respectively, for otherwise if $\Pi_i$ did not intersect $\v{a}_1$, we could choose $\v{a}_2$ or $\v{a}_3$ in its place. In other words, if a plane $\Pi_i$ cuts the $\v{a}_1$ axis at $i\frac{\v{a}_1}{h}$, the ``next plane" will cut the $\v{a}_1$ axis at $(i+1) \frac{\v{a}_1}{h}$. Moreover,  if $\v{r}_1$ and $\v{r}_2$ belong to $\Pi_i$ and $\Pi_{i+1}$, respectively, then 
        \begin{equation}
            (\v{r}_1 - i\frac{1}{h} \v{a}_1)\cdot \v{N} = 0
            \label{tproof1}
        \end{equation}
        and
        \begin{equation}
            (\v{r}_2 - (i+1)\frac{1}{h} \v{a}_1)\cdot    \v{N} = 0 
            \label{tproof2}
        \end{equation}
        The minimum distance between these two planes will be found by constructing the segment passing through the normal $\v{N} = h \v{b}_1 + k \v{b}_2 + l \v{b}_3$, which will imply that:
        \begin{equation}
            (\v{r}_2 - \v{r}_1)\cdot \frac{\v{N}}{\norm{\v{N}}} = \norm{\v{r}_2 - \v{r}_1} = T     
            \label{tprooft}
        \end{equation}
        If we now subtract equation \ref{tproof1} from equation \ref{tproof2}, we get:
        \begin{equation}
            (\v{r}_2 - \v{r}_1)\cdot \v{N}  = \frac{\v{a}_1}{h}\cdot \v{N}
            \label{tproof12}
        \end{equation}
        If we now replace $(\v{r}_2 - \v{r}_1)\cdot \v{N}$ of  equation \ref{tprooft} by the right hand side \ref{tproof12}, we get: 
        \begin{equation}
            \frac{\v{a}_1}{h}\cdot \frac{\v{N}}{\norm{\v{N}}} = T     
            \label{tprooff}
        \end{equation}
        or
        \begin{equation}
            \frac{2\pi}{\norm{\v{N}}} = T     
            \label{tprooff1}
        \end{equation}
        where we have used the fact that $\v{a}_1 \cdot \v{b}_{1}
        = 2\pi$ and $\v{a}_1 \cdot \v{b}_{2} = \v{a}_1 \cdot \v{b}_{3} = 0$.
        Note that over a period $T$ there is a total of one plane that is find (half of $\Pi_i$ and half of $\Pi_{i+1}$). The ``plane frequency" $f$ would then be $1/T$ and the ``angular plane frequency" would be computed as $2\pi/T$ = $\norm{\v{N}}$. The latter means that the norm of the normal vector $\v{N}$ to an (hkl) Miller plane gives us the ``angular plane frequency" in the $\v{N}$ direction .
        
    \end{appendix}
    
    \section*{Acknowledgements}
    
    This work was supported by the Laboratory Directed Research and Development program of Los Alamos National Laboratory under project number 20220431ER. This research used resources provided by the Los Alamos National Laboratory Institutional Computing Program. E.M. acknowledge support from Clemson University startup funds. E.M. was supported in part by the National Science Foundation EPSCoR Program
    under NSF Award \# OIA-1655740. Los Alamos National Laboratory is operated by Triad National Security, LLC, for the National Nuclear Security Administration of U.S. Department of Energy (Contract No. 89233218CNA000001). Additionally, we thank the CCS-7 group and Darwin cluster at Los Alamos National Laboratory for computational resources. Darwin is funded by the Computational Systems and Software Environments (CSSE) subprogram of LANL’s ASC program (NNSA/DOE). LA-UR-22-29886.
    
    \section*{References}
    
    \bibliographystyle{iopart-num.bst}
    \bibliography{references}

\end{document}